\documentclass[aps,tightenlines,preprint]{revtex4-1}
\usepackage{graphics,psfrag,amsmath}
\usepackage{epic,eepic}
\usepackage{dcolumn}
\usepackage{bm}
\usepackage{lmodern}
\usepackage{color} 
\usepackage{epsfig}
\usepackage{latexsym}
\usepackage{pstricks}
\newcommand{\be}{\begin{equation}}
\newcommand{\ee}{\end{equation}}
\newcommand{\bear}{\begin{eqnarray}}
\newcommand{\eear}{\end{eqnarray}}
\begin{document}

\title{Delbr{\"u}ck scattering in small scattering angle region}
\author{Taekoon Lee}
\email{tlee@kunsan.ac.kr}
\affiliation{Department of 
Physics, Kunsan National University, Kunsan 54150, Korea}


\begin{abstract}

The scattering of  photons of x-ray energy off a Coulomb field in very forward scattering
region may be thought as the refraction effect due to the Coulomb field. The cross section
of the scattering can be computed from the photon bending angle obtained in geometrical optics.
Its dependence on the fine structure constant is nonanalytic and 
softer than would the box diagrams of Delbr{\"u}ck scattering suggest.
The refractive scattering is a nonlocal effect occurring over the distance scale of the 
impact parameter, and therefore may not be described by the single-box diagrams. 
We suggest that the scattering cross section is a consequence of 
the sum of the asymptotic series of the multi-box diagrams.

\end{abstract}  \pacs{}  

\maketitle

High energy photons passing by the Coulomb field of a nucleus can deflect via the
vacuum polarization effect of the photons with the Coulomb field. This scattering of photons off the
Coulomb field is the Delbr{\"u}ck scattering (See for a review \cite{schumacher}), which has been observed 
using 2.754 MeV photons on lead nuclei \cite{schumacher1}. 
The scattering cross section has been computed with the one-loop
box diagrams \cite{cheng1,cheng2,
   cheng3,tollis,zhu,mork,koga}.

In this note we point out that in  very forward, small scattering angle region there is a
scattering mode that is nonlocal,
 occurring over the distance scale of the impact parameter of the photon. 
 Imagine a photon moving with a large impact parameter in a Coulomb field, with the photon wavelength
 much smaller than the impact parameter but larger than the electron Compton length.
Because the photons moving in  background electromagnetic fields behave as if they are traveling in a dielectric
medium with field-dependent refractive index, the deflection of the photons can be described by 
geometrical optics. In the Coulomb field the refractive bending of a light beam occurs over the
distance scale of the impact parameter, and its bending angle
has been calculated \cite{lee1}. Clearly, this is a nonlocal effect and therefore could not be described by the
box diagrams which have a distance scale of the electron Compton length.
The resulting scattering cross section of the refractive bending is nonanalytic in the fine structure
constant, and also has much softer dependence on it than the box diagrams would suggest,
 rendering the refractive cross section
be dominant over that of the box diagrams.
The refractive bending in the background field can be described by the Euler-Heisenberg effective interaction,
which we briefly review below.

The Euler-Heisenberg nonlinear electrodynamics induced by the
vacuum polarization effect in the quantum electrodynamics
 is described by the effective
Lagrangian which, in leading order is given by the box-diagrams, is
\bear
{\cal L}= -\frac{1}{4} F_{\mu\nu}F^{\mu\nu} +
\frac{\alpha^2\hbar^3}{90m^4c^{5}}
\left[( F_{\mu\nu}F^{\mu\nu})^2+\frac{7}{4}
( F_{\mu\nu}\tilde F^{\mu\nu})^2\right]\,,
\label{lagrangian}
\eear  
where $\tilde{F}_{\mu\nu}$ denotes the dual of the 
field strength tensor $F_{\mu\nu}$ \cite{euler,schwinger}.
 The Lagrangian is valid
in the low energy and weak field limit.

 For a 
photon  moving in a 
slowly varying background field,  with energy less than the 
electron rest mass, the background field effect can be encapsulated
in the refractive index $n$:
\bear
n=1+\frac{a\alpha^2\hbar^3}{45m^4c^5}|\bm{\hat{k}}\times 
\bm{E}+\bm{\hat{k}}\times(\bm{\hat{k}}\times \bm{B})|^{2}\,,
\label{index}
\eear
where $\bm{\hat{k}}$ denotes the unit vector in
 the direction of the photon propagation,
 and $a$
 is the birefringence constant 
that is either 8 or 14, depending on the photon polarization 
\cite{bialynicka}.
An interesting application of this is the bending of light in a Coulomb
field.
 Because the field-dependent index of refraction 
becomes larger at  stronger field, the incoming photon bends toward
the charge, in a fashion reminiscent of
the gravitational bending in general relativity. 
The bending angle can be easily
 computed in geometrical optics \cite{lee1}. For a photon with the 
impact parameter $b$ in the Coulomb field of a nucleus
 of charge $Ze$ it is given by 
\bear{} 
\theta(b)=\frac{a Z^{2}\alpha^3}{160} 
\left(\frac{\lambdabar_e}{b}\right)^4\,, 
\label{bendingangle} 
\eear{}
 where $\alpha$ is the fine structure constant, and
 $\lambdabar_{e}=\hbar/m_{e}c$ is 
the reduced electron Compton length. 
The  bending occurs over the distance scale of the
impact parameter.

The impact 
parameter in  Eq.~(\ref{bendingangle}) cannot be arbitrarily small, 
putting a limit on the size of the bending angle. Requiring that the 
Euler-Heisenberg interaction be a small perturbation to the Maxwell 
theory places a constraint on the field strength \cite{bialynicka}: 
\bear{} 
\frac{2a\alpha^{2}\hbar^{3}}{45m_{e}^{4}c^{5}}
 |F_{\mu\nu}|^{2} \ll 1\,, 
\eear{}
 where $F_{\mu\nu}$ denotes the background  field strength. 
For the Coulomb field  
 \bear{} E(r)=\frac{Ze}{4\pi 
r^{2}} 
\label{coulombfield}
\eear{}
 the constraint requires  the radius  satisfy
  \bear{} r\gg 
\lambdabar_{e}\left(\frac{aZ^{2}\alpha^{3}}{90}\right)^{\frac{1}{4}}\,, 
\label{const1}
\eear{} 
where the fine structure constant is given by 
$\alpha=e^{2}/4\pi\hbar c$.
 Even for a large $Z$ the radius satisfying the constraint can be 
fairly small. For instance, at $Z=100$ 
\bear{} r\gg 0.14 \lambdabar_{e}\,. 
\label{const2}
\eear{} 
Also the requirement that the background field be slowly 
varying demands \cite{bialynicka}:
\[ 
|\partial_{\lambda}F_{\mu\nu}|\ll \frac{m_{e}c}{\hbar}
|F_{\mu\nu}|\,,\] 
which for the 
Coulomb field corresponds to 
\bear{} r\gg \lambdabar_{e}\,.{} \label{con1} 
\eear{}

There is another constraint coming from the requirement that 
the corrections 
to the Euler-Heisenberg interaction be small.
Recall that 
the Euler-Heisenberg interaction arises from the low energy limit 
($k/m\to 0$, where $k$ denotes the momenta of the photons)
of the box 
diagrams of four external  photon lines. In a background 
Coulomb field the effective interactions of more than four 
photon lines  contribute to the light bending as well.
A simple  dimensional analysis shows that these interactions,  
which arise from the
 box diagrams of arbitrary, even number of external 
 photon lines, give rise to 
 terms
in powers of 
\bear{}
\frac{\alpha\hbar^{3}}{m_{e}^{4}c^{5}}
 |F_{\mu\nu}|^{2}\,,{}
 \label{corr}
\eear{}
relative to the Euler-Heisenberg term.
Substituting  the Coulomb field (\ref{coulombfield}) 
into Eq.~(\ref{corr}) and requiring these
 corrections  be 
 small leads to: 
\bear{}
r\gg \sqrt{Z\alpha}\lambdabar_{e}\,.
\label{const3}
\eear{}
The combined constraints above on the radius put a
 limit on the impact parameter
in the bending angle  (\ref{bendingangle}).
For heavy nuclei with $Z\alpha\sim 1$ it requires
\bear{}
b\gg \lambdabar_{e}\,.{}
\label{b-constraint}
\eear{}

 For a large $Z$ and
   small impact parameter, say $Z=100$ and $b=10\lambdabar_{e}$,
    the bending angle is $3.4 \times 
10^{-8} $ rad. 

Let us now comment on the constraints on the photon wavelength $\lambda$.  
It is clear now that for the validity of the
bending angle (\ref{bendingangle}) we need: (i) the Lagrangian 
(\ref{lagrangian}) be valid, and (ii) the geometrical optics  be 
applicable. Because the local Euler-Heisenberg term in the Lagrangian
 arises from the expansion 
in powers of $k/m_{e}$ (in natural units where $\hbar=c=1$), 
where $k$ denotes the momenta of the photon
lines, of the box diagrams, the photons bending off the Coulomb field
have to be soft, that is,
\bear{}
\lambda \gg \lambdabar_{e}\,,
\eear{}
limiting the bending to a small angle. 
The requirement that the geometrical optics be applicable to the
Coulombic bending of a photon with wavelength $\lambda$ demands that
the wavelength be smaller than the impact parameter $b$. Thus, for
a beam of photons of wavelength $\lambda$ the impact parameter is bounded
by:
\bear 
b\gg \lambda{}\,.
\eear{}
Therefore, the photons we consider are of x-ray energies and the scattering is
in very forward region with scattering angle less than a few tens of nano rad.

Our main focus in this note is the scattering cross section 
associated with the light bending and its effect on the
Delbr{\"u}ck scattering.
Using the general relation between the scattering angle and
 cross section \cite{goldstein}
 we obtain the cross section from the
bending angle (\ref{bendingangle}):
\bear{} 
\frac{d\sigma}{d\Omega} 
=\frac{b}{\theta}\left|\frac{db}{d\theta}\right| 
=\frac{1}{4}\sqrt{\frac{a Z^2\alpha^3}{160}} 
\frac{\lambdabar_e^2}{\theta^{\frac{5}{2}}}\,,
\label{cross}
\eear 
which should be  valid for very small scattering angles that are, as noted, at
 $Z=100, b=10\lambdabar_{e}$,  bounded by 34 nano rad.

A remarkable feature of this result is that 
the cross section is nonanalytic in the fine structure constant. The origin of the nonanalyticity
can be easily understood by a dimensional analysis.  Because the cross section is achromatic,
that is, independent of the photon energy, 
the particular form of the
 nonanalyticity 
 simply arises from the fact that the only dimensional parameter 
 in the Lagrangian (\ref{lagrangian}), the coefficient of
 the Euler-Heisenberg  term, is of mass dimension -4, which gives
 rise to the square root in Eq.~(\ref{cross}),  with the additional factor
   $Z^{2}\alpha$ within it
    coming from the Coulomb field.

\begin{figure}[tb] \begin{center} 
\includegraphics[scale=0.43]{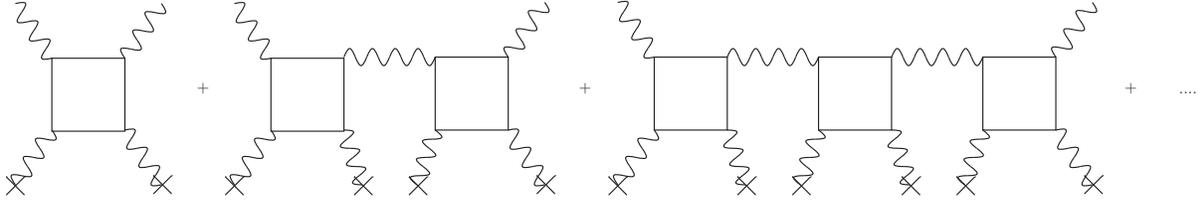} 
\end{center} 
\caption{Diagrams contributing to the light bending, 
where the lines crossed denote the Coulomb field.}
 \label{fig1} 
\end{figure}

This nonanalyticity shows that
the cross section cannot be obtained from the sum of Feynman 
diagrams (the box diagrams) to a finite order. Microscopically,
 the diagrams that contribute 
to the light bending would be the
multi-box diagrams like those
in Fig.~\ref{fig1}, since the Euler-Heisenberg interaction arises
from the single-box diagrams, and the arbitrary number of links of the
box diagrams as in Fig.~\ref{fig1}
 is in accordance with the nonlocality of the refractive scattering.
 The diagrams, 
a series in  powers of $Z^{2}\alpha^{3}$, 
  should be an asymptotic series (a generic feature 
  of perturbative amplitudes in field theory) since,
 otherwise, the sum of the diagrams would
 be analytic in the expansion parameter.
 These considerations suggests
 that the cross section (\ref{cross}) be the
   sum of the asymptotic
series of the multi-box diagrams to all orders. 
Assuming this conjecture be true, 
 it is remarkable
that the sum has such softer dependence on $Z^{2}\alpha^{3}$, hence
a much larger cross section, than
would each diagrams suggest. 

The multi-box diagrams have not been 
  considered in
  studies of the Delbr{\"u}ck scattering
   \cite{cheng1,cheng2,
   cheng3,tollis,zhu,mork,koga}.
  So far considered are only the single-box diagrams, 
  those with a single electron loop with
   four or more  even number of external photon lines.
  The amplitudes of those  diagrams  have
  $Z$-dependence of $Z^{2}\alpha^{3}$ in the
  leading order  and
  corrections to it in powers of $Z^{2}\alpha^{2}$, whereas 
  those of the multi-box diagrams
  are in powers of $Z^{2}\alpha^{3}$. 
  Thus it may appear at a given power of $Z$ the amplitudes of
  the multi-box diagrams are suppressed 
  by a factor $\alpha$ compared to those of the
  single-box diagrams.
  However, the cross section from the light bending suggests 
  this naive power counting as well as 
  the cross section from  the perturbative diagrams
   could be misleading. 
   The cross section from the 
   box diagrams at leading order, for instance, 
   has $Z$-dependence of $Z^{4}\alpha^{6}$, 
   which is completely different
    from that of (\ref{cross}).  Clearly, for nuclei with
   $Z\alpha\sim 1$ and in small $\alpha$  limit, 
   the cross section   from the
   light bending is dominant over any finite 
   order perturbative cross section.

  Furthermore, the cross section (\ref{cross}) is achromatic, 
  whereas the scattering 
  amplitude from the leading order box diagrams vanishes 
  as $\omega^{2}/m^{2}$ in $\omega/m\to0$ limit, 
  where 
 $\omega$ is the photon energy \cite{mork}. This implies that 
 for a soft photon
 the  cross section  from  the light bending should be dominant over
  the perturbative cross section in the very forward region.
  
  Thus the  cross section of the box diagrams is doubly
  suppressed at low energies 
  compared to that of the light bending:
  by energy dependence and $Z,\alpha$-dependence. Therefore,
  we would expect the  light bending should play an important role
  in the Delbr{\"u}ck scattering of soft photons (x-rays with 
  $\omega$ less than
  electron mass) in very forward scattering 
  region ($\theta\leq \text{ a few tens of nano rad}$), and potentially of 
  high energy photons as well. It would be interesting to see
  whether the effect can be studied  with the measurement of 
  the refractive index of Silicon or other metals \cite{habs,naito}.

\begin{acknowledgments}
I am thankful to S. Han for encouragement. 
This research was supported by Basic Science 
Research Program through the National 
Research Foundation of Korea
(NRF), funded by the Ministry of 
Education, Science, and Technology
(2012R1A1A2044543).
\end{acknowledgments}

\bibliographystyle{apsrev4-1}
\bibliography{delbruck-lightbending}

\end{document}